# From ChatGPT, DALL-E 3 to Sora: How has Generative AI Changed Digital Humanities Research and Services?

Jiangfeng Liu[1,2]    Ziyi Wang[1,2]    Jing Xie[3,4]    Lei Pei[1,2*]

**Abstract**

Generative large-scale language models create the fifth paradigm of scientific research, organically combine data science and computational intelligence, transform the research paradigm of natural language processing and multimodal information processing, promote the new trend of AI-enabled social science research, and provide new ideas for digital humanities research and application.

This article profoundly explores the application of large-scale language models in digital humanities research, revealing their significant potential in ancient book protection, intelligent processing, and academic innovation. The article first outlines the importance of ancient book resources and the necessity of digital preservation, followed by a detailed introduction to developing large-scale language models, such as ChatGPT, and their applications in document management, content understanding, and cross-cultural research. Through specific cases, the article demonstrates how AI can assist in the organization, classification, and content generation of ancient books. Then, it explores the prospects of AI applications in artistic innovation and cultural heritage preservation. Finally, the article explores the challenges and opportunities in the interaction of technology, information, and society in the digital humanities triggered by AI technologies.

## 1 Introduction

As an essential part of the historical and cultural heritage, ancient book resources are the valuable spiritual wealth of the Chinese nation and a necessary carrier of the excellent traditional Chinese culture (Zhang, 2022), which contains a wealth of

---

[*] 1. Data Intelligence and Cross-Innovation Laboratory, Nanjing University, Nanjing 210023, China

2. School of Information Management, Nanjing University, Nanjing 210023, China

3. School of Health Economics and Management, Nanjing University of Chinese Medicine, Nanjing 210023, China

**Corresponding author:**

Lei Pei, Nanjing University, No.163 Xianlin Avenue, Qixia District, Nanjing, 210023, China. Email: plei@nju.edu.cn

philosophical ideas, cultural essence, and historical information. The research and protection of ancient books are of great significance for the inheritance of Chinese culture, the enhancement of cultural self-confidence, and the excavation of the treasures of traditional civilization. However, due to the limitations of age and preservation conditions, the protection of ancient books faces many difficulties. The material damage of ancient books is unavoidable, and it is difficult to protect the original originality of ancient books. As a regenerative conservation measure, digital antiquities publishing can reproduce the spiritual content of antiquities and disseminate antiquities through social publishing (Gu, 2020).

The paradigm of scientific research determines the breadth and depth of humanity's exploration of unknown fields, and the development of global science is entering the fifth paradigm. Different from the data paradigm (fourth paradigm), which relies on massive data to analyze the intrinsic mechanistic relationship of affairs, the fifth paradigm further emphasizes the integration of data and scientific mechanism, combining data science and computational intelligence, introducing intelligent technology and strengthening the reasoning link (Zhang et al., 2023). AI has made significant progress in the past decades thanks to improved machine learning, big data, and computational power(Radanliev, 2024). Large-scale language models originating from AI have accelerated this change in the scientific research paradigm. In recent years, the advances in deep neural networks, natural language processing techniques, and large-scale language models have enabled the digitization and intelligent processing of ancient resources to show excellent performance and promising broad application prospects. This paper discusses the opportunities and challenges of large-scale language models from the research fields and computational and digital humanities application scenarios.

## 2 Review of large language models

The development of large-scale language models goes through the process from simple to complex, from single modality to multimodal, marking the in-depth exploration of artificial intelligence in natural language processing and machine learning and reflecting the continuous deepening of human understanding of language and cognitive processes. The development of large-scale language models can be divided into three stages.

In the primary stage, in 2017, Vaswani et al.(Vaswani et al., 2017) proposed the Transformer architecture and made a breakthrough in machine translation tasks. In 2018, Google proposed the BERT (Bidirectional Encoder Representation from Transformers) model(Devlin et al., 2019), which improves the depth of text representation using bi-directional training, especially makes a breakthrough in the ability to understand the context and is widely used in tasks such as text categorization, named entity recognition, and automated Q&A. The same year, OpenAI proposed the GPT (Generative Pre-trained Transformer)(Radford et al., n.d.) model based on the stacked Decoder structure in the Transformer architecture.BERT and GPT opened the era of pre-trained language models with "unsupervised pre-training" + "pre-trained language model." The "unsupervised pre-training" + "supervised fine-tuning" paradigm has gradually become the research trend in natural language processing. 2019, OpenAI released the GPT-

2(Solaiman et al., 2019) model with 1.5 billion parameters, and Google released the T5(Raffel et al., 2023) model with 11 billion parameter sizes. In 2020, OpenAI released the GPT3 model with 175 billion parameters.

Due to the large number of parameters in Large Language Models, performing task-specific domain fine-tuning on them is incredibly costly. Thus, researchers have begun to explore how to fully utilize the potential of Large Models without performing single-task fine-tuning scenarios. In the GPT-3 model, a few-shot learning approach called Context Learning(Brown et al., 2020) was proposed, which splices a small number of annotated samples from different tasks into the text-to-be-analyzed front-end input model for reference without modifying the model parameters. However, its performance still needs to improve compared to supervised fine-tuning. In 2022, the Instruction Fine-Tuning(H. W. Chung et al., 2022) scheme was proposed, which converts many different classes of NLP. This program converts many NLP tasks of different categories into a generative natural language understanding-type framework that builds "question-answer pairs" for training. In the same year, InstructGPT(Ouyang et al., 2022), based on "supervised fine-tuning + reinforcement learning," was proposed, which uses a small amount of supervised data to make a large model understand human instructions.

Since the release of ChatGPT in November 2022, large-scale language modeling research has officially entered a breakthrough phase of rapid development. Users can interact with ChatGPT through dialogue boxes and achieve various functions such as automatic question and answer, code generation, mathematical computation, logical reasoning, etc. It performs well in open-domain questions and answers the multi-round questions and answers and other generative NLP problems. In March 2023, the upgraded version of GPT-4 was released, which achieves cross-modal comprehension capability and supports image and text input and outputs text. In the same month, the ChatGPT Plugin plugin feature was launched, and the plugin shop provides a variety of third-party plugins for users to use. In May, the ChatGPT APP is released. In July, the code interpreter Code Interpreter is released. The textual graph model DALL-E 3 was released in September, and Microsoft began integrating the GPT-4-based Copilot functionality. In November, the GPT-4 corpus was updated to 04/2023; new gpt-4 all tool mode was added to incorporate all plug-in capabilities; new multimodal functionality was added to the development platform, including vision, image creation (DALL-E 3), and text-to-speech (TTS); and the GPTs functionality was released, which allows users to customize GPTs and publish them in the GPT shop. In January 2024, the gpt-4-0124-preview model was released with more robust performance, lower tariffs, and updated versions of gpt-3.5-turbo, text-embedding-3. On 15 February, OpenAI released the first video generation model - -Sora, which inherits the picture quality of the DALL-E 3 model and can generate 1-minute HD videos according to instructions.

Sora is a revolutionary text-to-video model OpenAI developed, representing a significant breakthrough in artificial intelligence. The model can generate video based on descriptive cues, expand existing video forward or backward, generate video from still images, and more. As of March 2024, Sora has yet to be released for public use, but its technical demonstrations and potential applications have attracted widespread attention. The team chose "Sora" (Japanese for "sky") as the project name to symbolize its "unlimited creative potential." Sora is one of many models in the history of technology that attempt to convert text to video. Still, it uses a diffusion transformer mechanism similar to that of DALL-E 3, which generates video by denoising 3D "patches" in latent space and then transforming them into standard space. Video. The advantage of this scheme is that it not only augments the training data with a video-to-text model but also demonstrates how to learn from the dataset to create 3D graphics and automatically generate different videos.

Sora has far-reaching implications for research and industry development in fields such as the humanities and the arts, providing new tools for relevant research and enabling researchers to understand and present historical events and cultural practices more intuitively. In content creation industries, such as film, television, advertising, and gaming, Sora offers a new way to generate high-quality video content quickly and inexpensively, potentially transforming the content creation process and industry model but raising concerns about the existing job market and copyright issues for AI products. Sora also raises concerns about the authenticity of online information, particularly in terms of political propaganda and the dissemination of disinformation. Sora is also a source of concern about the authenticity of online information, particularly regarding political propaganda and disinformation. While it offers new possibilities for artistic creation and cultural communication applications, the risk of misuse of its capabilities must be addressed. It is, therefore, essential to regulate and limit this emerging technology to ensure that it is used responsibly for healthy economic, cultural, social, and technological development.

## 3 Generative AI Transforms Digital Humanities Research

According to Popper's three-world theory, the third world is the world of objective knowledge, and human society has formed a particular order of knowledge production over thousands of years of development(Gao, 2023). Traditional knowledge formation takes a long time, both in terms of theoretical exploration and experimental research, while the large-scale language model represented by ChatGPT achieves light-speed knowledge production. With the technology iteration, the quality of knowledge production is also greatly improved, and the phenomenon of large model illusion is significantly reduced. As a non-decision-making AI, large-scale language models have instrumental value for research in humanities and social sciences(Chen, 2023).

### 3.1 Knowledge Organisation and Documentation Management

The secondary disciplines of information resource management, digital humanities, antiquities preservation, and bibliography have similarities with library and intelligence in the research paradigm, aiming at effective organization, storage, retrieval, and utilization of information resources to promote the sequencing of knowledge and optimize knowledge services. Ancient literature is the core research object of digital humanities research, carrying rich historical and cultural information. The effective management of ancient literature is significant for protecting and inheriting excellent cultural heritage.

The introduction of generative AI makes it possible to process documents on a large scale and understand the content and structure of complex documents. It can automatically process text and assist in the organization, cataloging, and classification of ancient documents, thus improving their accessibility and utilization effectiveness (Figure 1). AI can learn the linguistic characteristics and structural patterns of ancient papers and automatically identify the subject matter, historical period, and other information of the documents, thus improving the efficiency of the document management system. Efficiency of the management system. In addition, AI can analyze the content and semantics of digital humanities research documents to help build a more precise and detailed classification system, improve the accuracy of the retrieval system and user experience, and assist in the construction of mind maps, which can show the internal structure of the research and the knowledge connection in a visual way.

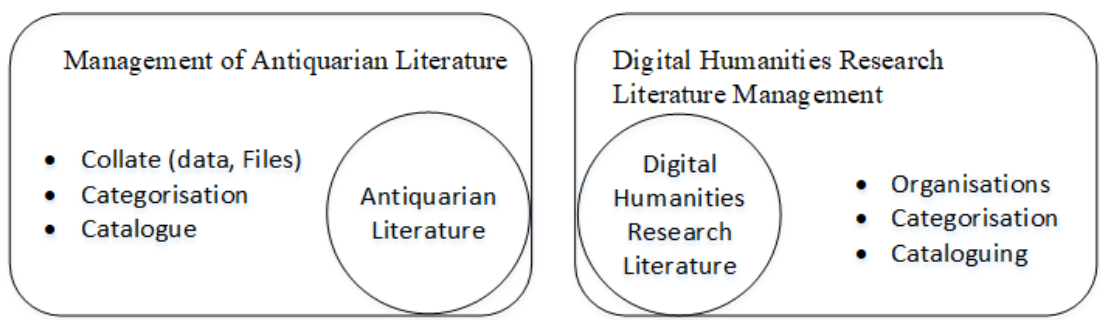

Figure 1 Knowledge Organisation & Documentation Management

## 3.2 Content Understanding and Knowledge Production of Ancient Literature

Complete ancient book protection and digital humanities research should include the original protection and digitization of ancient books, intelligent processing of texts, humanities computational analysis, and other processes. The primary data of the research comes from ancient literature, and some ancient texts are often mutilated due to historical reasons. ChatGPT can predict and generate the possible contents of the missing texts based on the existing literature content and context. For example, analyzing the writing style and context of the document and the author's historical background provides one or more options for text completion, thus assisting humanities scholars in restoring or reconstructing the document's original appearance. DALL-E 3, conversely, can regenerate or recover the content of documents containing illustrations or diagrams based on textual descriptions to improve the understanding and presentation of the documentation (see Figure 2).

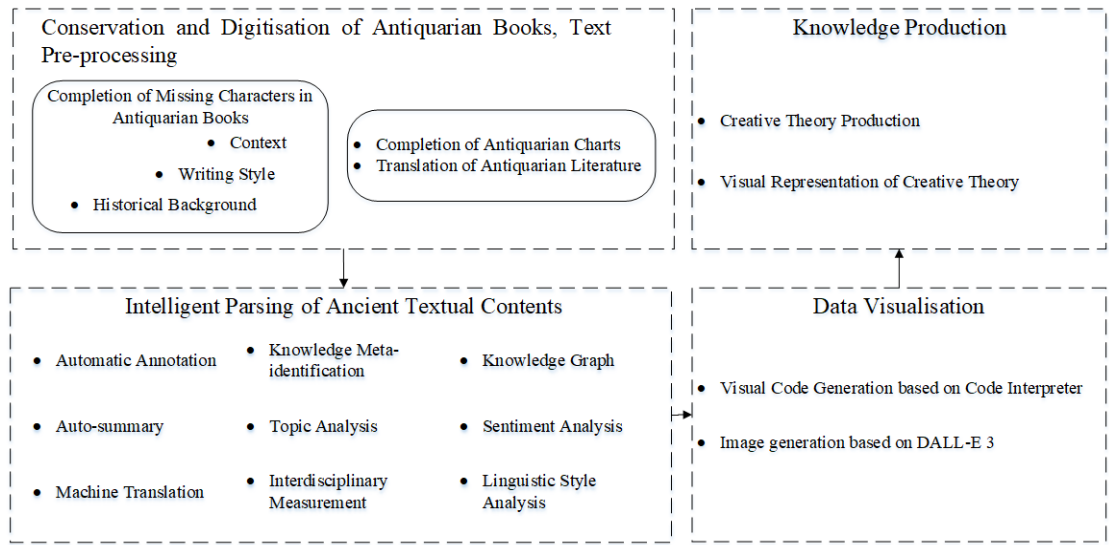

Figure 2 Literature content understanding and knowledge production

For humanities scholars in cross-cultural studies, the language barrier constitutes a significant challenge in their analysis of ancient texts from different regions. Large Language Models such as ChatGPT can translate literature with high quality, and even for rare languages, cross-cultural translation can be achieved by its vast training corpus(Ross, 2023). This expands the research horizons of humanities scholars, makes cross-cultural and even cross-temporal academic research more feasible, and promotes cross-cultural sharing and exchange of documentary resources.

For traditional humanities scholars, processing a large amount of text data and extracting critical information from it is complicated and trivial. They need help to use high-level programming languages, natural language processing techniques, and visual analysis techniques for automatic processing and visual analysis of large amounts of text. In contrast, large-scale language models can be important in this process. In terms of text content parsing, the use of appropriate commands can directly guide the large-scale model to parse historical documents(González-Gallardo et al., 2023), identify the named entities and relationships of ancient books from the text, construct a knowledge map of ancient books, identify the elements of ancient book events, identify the citation sentence, conduct text sentiment analysis, and achieve automatic summary, automatic annotation, theme and keyword generation of ancient books. For textual data visualization, ChatGPT's code interpreter(OpenAI, n.d.) function can generate code for constructing visual analysis charts, for example, for identifying ancient Chinese medicine prescriptions and case data analysis(Li & Zheng, 2023). Tools like DALL-E 3 and Sora can generate various data visualization images, including timelines, maps, and character relationship diagrams, such as realistic cave murals and contemporary art style works (Levine & Hausman, 2024). The introduction of big models has cracked the technical barriers between text processing and visualization analysis, providing humanities scholars with essential tools for digital humanities research and making relevant research conclusions more straightforward to understand, thus promoting academic communication and knowledge dissemination.

The expression of digital humanities research results lies in realizing the creative reproduction of humanities knowledge. When exploring new research themes or attempting to interpret existing materials from new perspectives, big models can be used to propose new research questions, hypotheses, or theories based on existing research materials and data, stimulating new academic thinking and providing new ideas for research. DALL-E 3 and Sora, on the other hand, can take such creative concepts and explore and express them through visual artworks, providing new ways of expression and dimensions of thinking for academic research.

The preceding discussion suggests that generative artificial intelligence is vital in promoting literature content understanding and creative knowledge production. The application of large-scale language models not only improves the efficiency of digital humanities research but also expands digital humanities research methods, broadens the horizons of digital humanities scholars, injects new vitality into creative knowledge production, and demonstrates great potential in promoting academic innovation and communication.

### 3.3 Humanistic Computing and Artistic Innovation

At the intersection of the arts and humanities, generative AI technologies are opening up new research areas and demonstrating their unique appeal. Multimodal art exploration, stylistic discernment, and cultural insight, the revival of cultural heritage in digital form, and the study of historical sound reproduction are possible research directions in this field, which, although focusing on different centers of gravity, together depict a technologically innovation-driven, multimodal and interconnected research ecology, and jointly promote the in-depth understanding and innovative expression of art and culture.

Research on art creation using deep learning techniques has a long history. Existing studies include the use of a Generative Adversarial Network (GAN) (Nie & Pu, 2023; Sun et al., 2019; Wang et al., 2023) to generate Chinese landscape-style paintings(Way et al., 2023), ink paintings(Wu et al., 2023), and flowers(Fu et al., 2021) and to convert traditional Chinese paintings into realistic images(C.-Y. Chung & Huang, 2023), etc. Large Language Models provide new ideas for optimizing and improving artistic creations and discriminating against them. Based on the multimodal Large Language Model, artistic exploration, and stylistic discernment, cultural insight constitutes the cornerstones of generative AI application in artistic understanding and innovation. The former focuses on the use of generative AI technology to understand and generate artistic thinking in multi-media forms, including images and texts, to promote the innovation and development of artistic expressions and to provide artists with new tools and new ideas for their creations.AI paintings can display human subjective consciousness and aesthetic attributes, and by constructing a method for creating assistant paintings that combines the Large Language Model with the Text-Image Generation Model, it can be used as a tool for the creative process. By constructing an assistant painting creation method, combining a large language model with a "text-image generation model," AI painting can provide precise content control for the generation of paintings(Lu et al., 2023). The latter focuses on the cross-modal simulation of different art styles, such as transforming novels into videos, to help researchers understand different art genres more intuitively and provide new perspectives for analyzing art styles to gain insights into the cultural connotations embedded in them. The organic integration of the two demonstrates the critical role of generative AI technology in promoting artistic understanding and creative expression.

In the digital age, the inheritance and protection of cultural heritage face unprecedented opportunities. Technologies such as big data, artificial intelligence, and deep learning can not only help repair and preserve damaged images of cultural heritage but also create new visual works based on ancient documentary materials, and the revival of cultural heritage based on digital forms provides people with a new perspective and a new solution for experiencing and understanding the history and culture of human beings. (i) Image restoration of cultural heritage usually involves restoring images of ancient documents, ancient artifacts, and architecture(Basu et al., 2023; Kumar et al., 2024). Well-established practice is to restore the integrity of an image by recognizing damaged parts of an image and filling in the missing parts with the help of deep learning techniques, especially convolutional neural networks (CNNs)(Liu Yixuan et al., 2023) and generative adversarial networks (GANs)(Jin et al., 2020), and inferring the possible appearance of the damaged parts based on the content of the surrounding image. In addition, by combining historical documents and interdisciplinary research results, it is also possible to more accurately reproduce the original artistic style and details so that the restoration work is not only limited to restoring the physical form but also preserves the historical and artistic value of cultural heritage as much as possible. (ii) Large Language Models and Multi-Modal Models can transform descriptions in ancient documents into concrete visual images, reproducing historical events or recreating ancient scenes for creating cultural and artistic works. This transformation from text to image provides a new way of learning and experiencing history and a source of inspiration for artistic creation. In particular, style migration research can apply ancient art styles to modern designs, allowing ancient cultural elements to be presented to the audience in a new art form. Finally, combining Augmented Reality (AR) and Virtual Reality (VR) technologies based on ancient documents and archaeological excavations creates an immersive historical experience space, allowing users to experience historical scenes and cultural stories as if traveling through time and space.

Speech generation research has a long history(Kaur & Singh, 2023). Historical sound reproduction refers to the use of AI to explore the construction of the sounds of ancient languages and music, which not only deepens the understanding of linguistic and musical cultural heritage and provides new directions for linguistic and artistic research but also further promotes cross-cultural communication and understanding. Its development trajectory can be traced back to the mid-20th century when researchers first attempted to simulate human speech with the help of electronic synthesizers. Although the early results sounded rough and lacked natural rhythm, they marked the beginning of the journey of exploring speech reproduction with the aid of technology. With the passage of time and technological innovations, especially in the computer age, speech synthesis has undergone a profound transformation from simple simulation to high intelligence.

As an important application of this branch of research, the recreation of historical sounds focuses on the powerful potential of AI technology in building and restoring ancient languages, music, and other acoustic heritage. This process goes beyond the digital repair or restoration of past audio recordings and delves into the scientific reconstruction of lost or severely degraded speech systems. It involves the comprehensive analysis of ancient documents, archaeological evidence, linguistic kinship, music theory, historical customs, and other multifaceted information, combined with modern phonetics, acoustics, computer science, and artificial intelligence algorithms to construct highly realistic models of ancient speech. These models are not only capable of recovering the pronunciation characteristics of the language in a specific historical period or cultural context, such as unique combinations of vowels and consonants, pitch changes, and intonation patterns but also reproduce the timbre of specific musical instruments, the performance style of ancient musical works and even the environmental acoustics of particular occasions.

The importance of this work is reflected on several levels. Firstly, it greatly enriches linguistics and art's research tools and contents. Through the auditory reconstruction of ancient languages, scholars can understand more intuitively the laws of phonological evolution, dialectal differences, and the impact of cultural exchange, opening up new experimental paths for research in linguistic history, dialectology, and phonetics. Regarding art, reproducing historical music sounds can help reveal the more profound connotation of ancient music aesthetics, performance practices, and social functions and promote the development of archaeology, music history, and related cross-disciplines. Secondly, the reproduction of historical sounds has far-reaching significance for the protection and inheritance of cultural heritage. It enables museums, archives, libraries, and other institutions to present historical sound resources in an interactive and immersive way so that the public can transcend time and space, listen to the sounds of ancient civilization, and feel its unique charm, thus enhancing the public's historical and cultural literacy and strengthening national identity and cultural confidence. Furthermore, this technology promotes cross-cultural dialogue and understanding. Through the reconstruction of ancient language and music by AI, people from different cultural backgrounds can cross the barriers of language and time, directly contact and perceive each other's distant ancestral voices, enhance their knowledge of their respective cultural origins, and bring each other closer, which is conducive to the enhancement of international cultural exchanges and harmonious coexistence in the context of globalization. In addition, historical sound reproduction also provides unprecedented creative materials for multimedia industries such as film and television, games, and virtual reality. These realistically reproduced ancient sounds can be used in historical works, giving them a higher sense of historical authenticity and artistic impact, enhancing the audience's experience of watching films or playing games, and also helping to develop more attractive historical teaching resources in education.

The research above areas not only promotes more profound understanding and innovation in art and culture but also points the way for future research in the history of humanities and art and the protection of cultural heritage, highlighting the critical role of generative AI technology and its potential in promoting the progress of the cultural industry.

# 4 Challenges posed by generative AI for the digital humanities

Generative AI not only presents opportunities for digital humanities research, education, and knowledge services but also many challenges(See Figure 3).

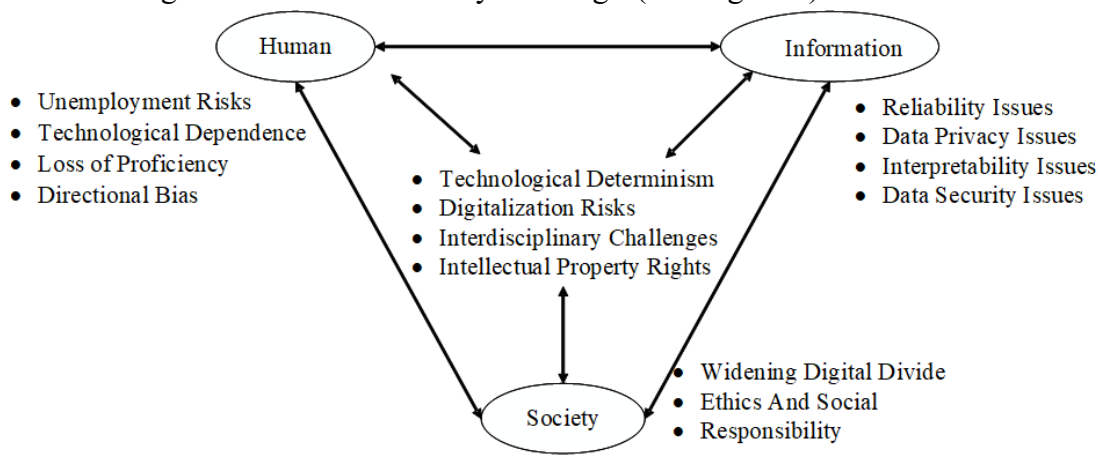

Figure 3 Challenges Generative AI Poses for the Digital Humanities Field

## 4.1 Technology & Human

From a humanistic perspective, the development and application of AI technology have brought varying degrees of convenience and efficiency gains to people's lives and work. However, the uncontrollability and complexity of AI itself require us to take a dialectical approach to the impact of technology on people. This subsection explores the possible negative consequences of rapidly developing AI technologies on people and focuses on four.

The first is the risk of skill replacement and unemployment due to AI. Changes in socio-economic structure due to AI technology are now a reality. While the development of AI technology also creates new industries and jobs, the more prevalent effect is now the replacement or even disruptive elimination of jobs. Such shifts may result in workers in specific sectors needing to retrain or switch careers to adapt to the new job market(Guliyev et al., 2023). The current technological innovation represented by artificial intelligence has already created an apparent wave of technological unemployment. On the one hand, the speedy update rate of digital technology far exceeds the update rate of organizations and workers' skills, making it difficult for the labor force to promptly adapt to the new requirements and environment brought about by new technologies. On the other hand, automation technology and AI technology can significantly reduce the demand for workforce in enterprises, thus saving costs, which has led to the market economy environment; the use of intelligence instead of manual labor has become the majority of enterprises' choice(Lima et al., 2021). Furthermore, AI systems can perform repetitive and standardized tasks quickly and accurately, which is usually more efficient than manual labor. This AI-optimised replacement of human labor is more common, especially in positions requiring less initiative and creativity from the workforce. The second is that AI technology may also increase people's technological dependence. As AI technology becomes more sophisticated, its performance and cost are gradually optimized, and it is embedded in daily work-labor scenarios with increasing frequency, significantly improving productivity. However, over-reliance on technology may weaken people's basic abilities, such as critical thinking, creative thinking, etc. In digital humanities and knowledge services, this may lead to a decline in the ability to understand raw materials and data in depth, which may be manifested in the over-pursuit of the visualization of data presentation at the expense of deeper excavation of the data itself.

In addition, AI may also lead to a loss of skills, and over-reliance on generative AI may lead to a degradation of the researcher's skills in traditional research skills and methodologies. Although the initial intention of introducing AI technology into various research and production environments was to assist manual processes in improving efficiency, with the continuous improvement of AI technology and the deepening of its integration with human production activities, AI technology has gradually moved closer to the dominant position. It has become more and more technologically dominant over research or production activities as a whole(Cabitza et al., 2023). Nowadays, the substitution of AI for human brain thinking ability has become increasingly apparent; for example, ChatGPT is often used by school students to assist in completing course assignments, which can improve course grades and learning efficiency in the short term. Still, in the long term, it will likely weaken the students' ability to think independently and solve problems creatively. The inappropriate use of AI technology may also lead to a bias in the direction of research, as the application of AI technology may lead to an overly biased research direction towards technologically achievable areas while neglecting research topics that require deeper thinking and human intuition. People's AI literacy tends to be uneven. Many people do not have sufficient screening and reflective ability, which will lead to a gradual tendency to be passive in the process of interacting with AI technology and an increasing reliance on automated systems, a process that tends to gradually erode people's motivation and enthusiasm for active thinking and innovative conceptualization, thus making research increasingly homogeneous and closed.

## 4.2 Technology & Information

From the information perspective, AI technology has undoubtedly brought a subversive boost to the research work; under the auspices of more powerful arithmetic power and more prosperous deep learning methods, the depth and breadth of the information processing work have been greatly improved. The massive amount of data that was difficult to deal with in the past is gradually being utilized. The depth of excavation that was previously untouchable has also steadily become possible. Still, there are some equally. However, in this process, some potential problems need attention.

Firstly, there needs to be more information quality and reliability. The wide application of AI technology has increased the speed of information generation and circulation. Still, at the same time, it has also brought about the problems of information overload and information quality control. Due to the possible inaccuracy of the original training data, coupled with the fact that complex model training may lead to overfitting phenomena and deficiencies in generalization ability, the accuracy of AI technology is often not sufficiently guaranteed. For example, some scholars organized dozens of professional physicians to assess the accuracy of GPT's answers to professional medical questions. The results showed that ChatGPT generated roughly accurate information for diverse medical queries. Still, the accuracy of the answers could not be adequately guaranteed, and its ability to deal with complex problems was limited(Johnson et al., 2023). The inaccurate information often misleads users to make wrong decisions and judgments, generating direct economic losses or other risks. In addition, the issue of data privacy and security also deserves to be taken seriously. With the popularity of IoT devices and the development of cloud computing, much personal and sensitive data is generated and processed. Users entrust sensitive information to IoT devices, and the security and privacy of this data are crucial when it is transferred to the cloud for storage and analysis. Finally, the information processing process may face interpretive and explainability issues, as generative AI may produce complex and insightful analyses. Still, generating these results is often "black-box" and challenging to explain and understand. Many AI models and intense neural networks process input data and make predictions through multiple layers of non-linear transformations. The inner workings of these models are often complex, making it difficult for even designers and developers to track and understand how the model derives a particular output from the input data(Toreini et al., 2019). This challenges digital humanities scholars who wish to gain insights through the research process. Several digital humanities projects are facing problems in several aspects mentioned above; for example, in some intelligent Q&A systems made for specific corpora and contexts, the system achieves the goal of providing accurate answers to questions about basic information about the content in the corpus. Still, the AI models often need to perform better regarding logically complex and analytical questions. Regardless of whether the answers are correct or incorrect, the user is often unable to verify the source and accuracy of the answer, which makes it always necessary to be wary of the results when using them.

## 4.3 Technology & Society

The impact of technology on social production is indisputable, and the convenience it brings to daily life is evident to all. However, while this process benefits society, it may exacerbate the digital divide, i.e., the uneven development of technology may exacerbate social inequality, especially in resource-poor regions. Differences in levels of economic development often lead to differences in the extent to which different people are exposed to AI technologies and related products. In contrast, differences in levels of education or personal technological literacy often lead to differences in the efficiency and ability of different people to use AI technologies. The interaction of all these possible economic factors, educational level factors, infrastructure, socio-cultural factors, and other elements ultimately exacerbates the digital divide. Its specific connotations can be understood in the following ways: (i) Access divide: This refers to the differences in access to ICT infrastructure (e.g., internet, computers, smartphones, etc.) between individuals or groups. For example, most households and individuals can access high-speed Internet in urban areas. In contrast, in rural or poor areas, people often need access to the same quality of Internet services due to inadequate infrastructure. (ii) Usage divide: Even where ICT can be accessed, there may be differences in the level of proficiency and frequency with which different individuals use these technologies. (iii) Skills divide: This relates to the level of skills individuals have when using ICT, including basic digital literacy, information retrieval, online communication, and advanced skills such as programming or data analysis. (iv) Outcome divide: Even when individuals can access and use ICT, the actual benefits they derive from it may differ(Aissaoui, 2022). For example, in some digital humanities projects that digitally store paper-based materials for the sake of preserving the original corpus, some projects may close the open access service to the original paper-based precursors after completing the construction of the digitized entry and presentation platform, which inadvertently makes it more difficult for people who have difficulty in using electronic technology, such as older people, to access the service.

The use of AI technology may also pose ethical and social responsibility challenges. Examples include issues of transparency in AI decision-making and algorithmic bias. In digital humanities research and knowledge services, it is an ongoing challenge to ensure that the application of technology follows ethical standards and respects human rights and social justice. Since the training of systems for generative AI relies heavily on the original training data, in the event of biases in the dataset, which may, for example, over-represent one group at the expense of others, the AI system may continue to learn and replicate these inequalities. Furthermore, even if the data themselves are neutral, if they reflect historical biases and social disparities, the AI system may inherit these biases. Due to differences in their cognitive structures, the designers of algorithms may also (consciously or unconsciously) encode personal biases or preconceptions into the algorithms. The criteria used to evaluate the performance of AI systems may likewise be biased, leading to system optimization that implies the wrong direction of tuning. The combination of these factors leads to the possibility of algorithmic biases in AI algorithms, such as employment discrimination, lending inequity, and miscarriage of justice(Henz, 2021). With the popularity of AI technology, the social responsibility issues faced by enterprises and research institutes have become increasingly prominent, and it has become necessary to justify and fully constrain the use of AI technology. For example, "digital life" has been widely discussed on the Internet, and many people believe that "resurrecting" people who have passed away digitally is disrespectful to life. The management and constraints on "digital lifeforms" are equally important. There is also a gap in managing and regulating "digital life."

## 4.4 Technology, Human & Society

When we look at the perspective of people and society, we can see that the use of AI technology promotes the possibility of better development of individuals at the social level. It must also be acknowledged that with the help of advanced computer technology, society can better integrate the results of the work of individuals to move forward; for example, more and more large-scale models have come into being through the joint efforts of countless researchers. These models are constantly being perfected during iteration, which is increasingly in line with the users' needs. These models are continually being improved and adapted to the needs of users, which significantly facilitates researchers and thus contributes to the progress and development of civilization as a whole. However, to ensure that AI technology plays a positive role, it is necessary to solve the many problems that may arise when combining AI with people and society.

The first thing that needs to be faced is the problem of academic ethics. However, it has become indisputable that using AI tools to assist academic research can save time and improve efficiency; the accompanying academic ethical issues should not be ignored. These issues are mainly reflected in (i) information bias and inaccuracy: AI models often have unavoidable bias and error in the training process, so its generation of content is often unable to ensure absolute equality and accuracy, and if the researchers include biased or error AI-generated content in their research results, it may affect the fairness and objectivity of scientific development. (ii) Intellectual property issues: AI-generated content is often generated based on the information in its massive training data, so if users directly reuse AI-generated content without distinguishing and pursuing the source, it may raise copyright issues. (iii) Academic integrity issues: If researchers borrow from AI-generated content without restriction and accelerate their research process with the help of AI tools in a large number of repetitions, they may violate the principle of academic integrity and the originality and innovation of their results will be questioned, or even judged as plagiarism(Lund et al., 2023).

In addition, copyright and intellectual property rights may likewise generate new risks due to the introduction of AI. Generative AI can be used to create texts, music, and other kinds of artistic works, and the attribution of copyright to these works has now become a complex and comprehensive issue. AI's problems in protecting intellectual property are manifold, one of which is the blurring of the boundaries between the subject and the object. In intellectual property law, the subject issue of AI involves whether it can be the subject of creation and the subject of the right regarding its eligibility(Liu, 2019). Based on the primary criterion of the legal level, only "human" can be regarded as the subject of intellectual property legal relations. Still, due to the excellent efficiency advantage of AI technology, AI has been getting closer and closer to the identity of the actual sense of the creative subject(Liang, 2017), which will lead to the non-subjectivity of AI as an object in the legal sense of the primary criterion has been challenged. The basic norm is challenged. The second is the subversion of labor theory; the traditional labor theory in interpreting artificial intelligence labor and its creations has apparent deficiencies. Artificial intelligence activities and traditional "labor" activities are significantly different, so artificial intelligence can not be regarded as the subject but also can not be the conventional sense of the object(Kop, 2019), which will bring more uncertainty to the attribution of rights and even legal issues related to this. According to the description, many kinds of AI art creation products are on the market, such as creating drawings, music, and videos. However, these AI creations are generated by the model in real-time according to the user's requirements and prompts, and many people believe that it is still essentially a "collocation" and "replica" of the model's original training data. Although these AI creations are generated in real-time by models based on user requests and prompts, many believe that they are still essentially "collages" and "replicas" of the original training data of the models, which complicates the copyright issues that may arise.

## 4.5 Overall risk of Technology at the Human, Information, and Social levels

In recent years, the outstanding performance of various types of AI models and their excellent performance in practical applications have led many people unconsciously towards technological determinism, i.e., the view that technological development is the main driving force for social change and cultural patterns and even that technology will be the only factor driving social progress. This view ignores the complex interactions between technology and society, which may lead to technological development that is detached from the needs of society, and at the same time, simplifies the complex relationship between technology and society, failing to take into account the role of human choice and decision-making in technological development. In addition to this, technological determinism may lead to pessimism and even undermine public participation and policy interventions. Therefore, we must recognize that technology has an essential impact on society. At the same time, we also need to see the role of society in shaping technology, i.e., technology and society interact with each other. This interaction constitutes the complexity of technological development(Hallström, 2022). To promote the better adaptation of AI technology to society, it is necessary to consider the needs and values of society in the design, development, and application of AI. Achieving sustainable development of AI technologies requires us to focus on how AI can be used for sustainability goals, the environmental impacts of AI development and use, and how to promote AI innovation without harming the environment, the economy, and society(Van Wynsberghe, 2021).

In the field of digital humanities, the use of generative AI for digitizing cultural heritage may be at risk of digitizing cultural heritage. Digitization allows cultural heritage to be appreciated and learned by a broader range of people and facilitates communication and understanding between different cultures. This helps to establish cross-cultural dialogue and increase the diversity and inclusiveness of the global community(Leshkevich & Motozhanets, 2022). However, there are still some problems with digitizing cultural heritage: (i) weakened interactivity and loss of value perception: Although digitization can provide users with an image or video experience that is as close as possible to that of the original object, it can never replace the original artifact's own material characteristics and aesthetic value, nor can it provide sufficient interactivity perception, which may affect the sense of value identity and emotional connection to cultural heritage(Hou et al., 2022). (ii) Technical dependence and sustainability issues: The digitization of cultural heritage relies heavily on long-term technical maintenance and data preservation, while long-term technical maintenance requires sufficiently stable financial support, which means that once the support for data preservation is interrupted or weakened, it may result in the loss of cultural heritage or damage to the problem, thus bringing risks to the inheritance of cultural heritage(Gervasi et al., 2022). (iii) Data security issues: Due to the insecurity of the network itself, digitized cultural heritage may be exposed to the risk of cyber-attacks and data leakage, which is also one of the factors to be considered in the process of digitizing cultural heritage. (iv) Ethical and cultural sensitivity issues: The digitization process may encounter sensitive issues related to specific cultural traditions and values, so it is essential to ensure that digitization efforts respect and protect the diversity and complexity of cultural heritage(Zhao et al., 2020).

Digital humanities and computational humanities research often require interdisciplinary collaboration, which requires participants to have profound professional knowledge and an understanding of techniques and methods in other fields. Constructing an effective interdisciplinary collaboration model that promotes communication and learning between researchers from different backgrounds is the key to achieving synergistic development of technology, people, information, and society. However, interdisciplinary collaboration also faces challenges: (i) Communication barriers: Experts from different disciplinary backgrounds may use different terminologies and conceptual frameworks, reducing communication efficiency between collaborators or causing ambiguity and understanding bias(ÓhÉigeartaigh et al., 2020). (ii) Methodological differences: Each discipline has unique research methods and practices, which may be incompatible or difficult to integrate. Collaborators must find ways to harmonize and integrate different methodologies to work effectively(Gefen et al., 2021). (iii) Ethics and compliance issues: Different disciplines may have different ethical guidelines and compliance requirements, and often, collaborators from other disciplines may not be familiar with these rules. (iv) Sustainability of long-term collaboration: As participants in interdisciplinary collaborations often come from different disciplinary backgrounds or research fields, their academic intersections tend to be fewer in number, and thus, sustaining such spanning collaborative relationships is also an issue to be considered(Stenseke, 2022).

In conclusion, as a core driver of digital humanities, computational humanities research, and knowledge services, technology brings many implications and challenges in its interactions with people, information, and society. In the face of these challenges, sustained technological innovation, ethical scrutiny, policy formulation, and education and training are needed to ensure that the healthy development of technology can facilitate the creation and dissemination of knowledge and serve the overall progress of society.

## 5 Conclusion

With the rapid development of AI technology, large-scale language models have become the core driving force of digital humanities research, bringing unprecedented research tools and methods to traditional humanities. These technologies have enhanced research efficiency, expanded research horizons, and injected new vigor into academic communication and knowledge innovation. However, the development of technology has also brought many challenges, such as career replacement, technological dependence, information quality control, and data security. In the future, we need to strengthen ethical scrutiny and policy formulation while promoting technological innovation to ensure that the healthy development of technology can effectively serve the overall progress of society and promote cultural heritage and innovation.

## Acknowledgments

I want to acknowledge that this research was made possible with support from the School of Information Management, Nanjing University.

## Declaration of Conflicting Interests

## Funding

This paper is one of the research results of the major project of Philosophy and Social Science Research in Colleges and Universities in Jiangsu Province titled "Construction and Application of Pre-trained Models for Traditional Chinese Medicine Classics" (Grant No. 2023SJZD084), and the project of Postgraduate Research & Practice Innovation Program of Jiangsu Province, "Research on transforming the organization and evaluation of knowledge of literature with generative artificial intelligence" (KYCX24_****).

## Data Accessibility

No data were used in this study.

## ORCID iD

Jiangfeng Liu https://orcid.org/0000-0001-7268-7313
Ziyi Wang https://orcid.org/0009-0000-1823-7629
Jing Xie https://orcid.org/0000-0002-7242-4924
Lei Pei https://orcid.org/0000-0003-4754-4112

## Supplemental Material

There is no supplemental material for this article.

## Author Biography

**Jiangfeng Liu** is a PhD student in the School of Information Management at Nanjing University. He is an information science researcher interested in studying the application of artificial intelligence technology and natural language processing technology in the social sciences, focusing on intelligent knowledge organization and evaluation of scholarly literature, preservation, and intelligent processing of ancient literature.

**Ziyi Wang** is a graduate student in the School of Information Management at Nanjing University. She is an information science researcher. Her research focuses on researchers' growth and digital humanities topics.

**Jing Xie** is an associate professor at the School of Health Economics and Management at Nanjing University of Chinese Medicine. Holding a PhD, he also serves as a Master's Supervisor. His research focuses on applying artificial intelligence and natural language processing in the social sciences, particularly in intelligent organization, and evaluating ancient literature.

**Lei Pei** is the Dean, Professor, and Doctoral Supervisor of the School of Information Management at Nanjing University. He is an information science researcher interested in studying policy quantification, digital humanities, and information economics.